\documentstyle[11pt]{article}
\makeatletter \@addtoreset{equation}{section} \makeatother

\newcommand{\noi}{\vspace{12pt}\noindent}
\newcommand{\beq}{\begin{equation}}
\newcommand{\eeq}{\end{equation}}
\newcommand{\bea}{\begin{eqnarray}}
\newcommand{\eea}{\end{eqnarray}}

\newcommand{\e}[1]{{(\ref{#1})}}
\newcommand{\eq}[1]{{eq.\ (\ref{#1})}}
\newcommand{\es}[2]{{(\ref{#1}) and (\ref{#2})}}
\newcommand{\eqs}[2]{{eqs.\ (\ref{#1}) and (\ref{#2})}}
\newcommand{\Ref}[1]{{Ref.~\cite{#1}}}
\newcommand{\mb}[1]{{\mbox{${#1}$}}}

\newcommand{\ie}{{${ i.e.\ }$}}

\newcommand{\cf}{{cf.\ }}
\newcommand{\wrt}{{with respect to }}

\newcommand{\lhs}{{left-hand side }}
\newcommand{\rhs}{{right-hand side }}

\newcommand{\eps}{\varepsilon^{}}
\newcommand{\cR}{{\cal R}}
\newcommand{\Deltaone}{\Delta_{1}^{}}
\newcommand{\DeltaE}{\Delta_{E}^{}}
\newcommand{\DeltaF}{\Delta_{F}^{}}
\newcommand{\DeltaFE}{\Delta_{F,E}^{}}
\newcommand{\DeltaFR}{\Delta_{F,\cR}^{}}
\newcommand{\DeltaR}{\Delta_{\cR}^{}}

\newcommand{\Deltarho}{\Delta_{\rho}^{}}
\newcommand{\nuF}{\nu_{F}^{}}
\newcommand{\nuR}{\nu_{\cR}^{}}

\newcommand{\nurho}{\nu_{\rho}^{}}
\newcommand{\nR}{n_{\cR}^{}}
\newcommand{\VR}{V_{\cR}^{}}
\newcommand{\WE}{W_{E}^{}}
\newcommand{\XE}{X_{E}^{}}

\newcommand{\str}{{\rm str}}
\newcommand{\sdet}{{\rm sdet}}

\newcommand{\Hf}{{1 \over 2}}

\newcommand{\Ih}{{i \over \hbar}}
\newcommand{\Hi}{{\hbar \over i}}

\newcommand{\deder}[1]{{ 
 {\stackrel{\raise.1ex\hbox{$\leftarrow$}}{\delta^r}   } 
\over {   \delta {#1}}  }}
\newcommand{\dedel}[1]{{ 
 {\stackrel{\lower.3ex \hbox{$\rightarrow$}}{\delta^l}   }
 \over {   \delta {#1}}  }}
\newcommand{\papar}[1]{{ 
 {\stackrel{\raise.1ex\hbox{$\leftarrow$}}{\partial^r}   } 
\over {   \partial {#1}}  }}
\newcommand{\papal}[1]{{ 
 {\stackrel{\lower.3ex \hbox{$\rightarrow$}}{\partial^l}   }
 \over {   \partial {#1}}  }}
\newcommand{\rpa}[1]{{ 
 \stackrel{\raise.1ex\hbox{$\leftarrow$}}{\partial^r_{#1}}   }}
\newcommand{\lpa}[1]{{ 
 \stackrel{\lower.3ex\hbox{$\rightarrow$}}{\partial^l_{#1}}  }}

\newcommand{\proofbox}{\begin{flushright}
${\,\lower0.9pt\vbox{\hrule \hbox{\vrule
height 0.2 cm \hskip 0.2 cm \vrule height 0.2 cm}\hrule}\,}$
\end{flushright}}

\newtheorem{theorem}{Theorem}[section]

\newtheorem{lemma}[theorem]{Lemma}

\newtheorem{proposition}[theorem]{Proposition}

\begin{document}
\thispagestyle{empty}
\title{\Large{\bf Odd Scalar Curvature in Field-Antifield Formalism}}
\author{{\sc Igor~A.~Batalin}$^{ab}$ and {\sc Klaus~Bering}$^{ac}$ \\~\\
$^{a}$The Niels Bohr Institute\\Blegdamsvej 17\\
DK-2100 Copenhagen\\Denmark \\~\\
$^{b}$I.E.~Tamm Theory Division\\
P.N.~Lebedev Physics Institute\\Russian Academy of Sciences\\
53 Leninisky Prospect\\Moscow 119991\\Russia\\~\\
$^{c}$Institute for Theoretical Physics \& Astrophysics\\
Masaryk University\\Kotl\'a\v{r}sk\'a 2\\CZ-611 37 Brno\\Czech Republic}
\maketitle
\vfill
\begin{abstract}
We consider the possibility of adding a Grassmann-odd function $\nu$ to the
odd Laplacian. Requiring the total $\Delta$ operator to be nilpotent leads to 
a differential condition for $\nu$, which is integrable. It turns out that
the odd function $\nu$ is not an independent geometric object, but is instead
completely specified by the antisymplectic structure $E$ and the density
$\rho$. The main impact of introducing the $\nu$ term is that it makes
compatibility relations between $E$ and $\rho$ obsolete. We give a geometric
interpretation of $\nu$ as (minus $1/8$ times) the odd scalar curvature of an
arbitrary antisymplectic, torsion-free and $\rho$-compatible connection. We
show that the total $\Delta$ operator is a $\rho$-dressed version of
Khudaverdian's $\Delta_E$ operator, which takes semidensities to
semidensities. We also show that the construction generalizes to the situation
where $\rho$ is replaced by a non-flat line bundle connection $F$. This
generalization is implemented by breaking the nilpotency of $\Delta$ with an
arbitrary Grassmann-even second-order operator source.
\end{abstract}
\vfill
\begin{quote}
PACS number(s): 02.40.-k; 03.65.Ca; 04.60.Gw; 11.10.-z; 11.10.Ef; 11.15.Bt. 
\hrule width 5.cm \vskip 2.mm \noindent 
$^{b}${\small E-mail:~{\tt batalin@lpi.ru}} \hspace{10mm}
$^{c}${\small E-mail:~{\tt bering@physics.muni.cz}} \\
\end{quote}

\newpage

\section{Introduction}
\label{secintro}

\noi
Conventionally \cite{schwarz93,bt93,kn93,hz94} the geometric arena for
quantization of Lagrangian theories in the field-antifield formalism
\cite{bv81,bv83,bv84} is taken to be an antisymplectic manifold \mb{(M;E)}
with a measure density \mb{\rho}. Each point in the manifold \mb{M} with local
coordinates \mb{\Gamma^{A}} and Grassmann parity
\mb{\eps_{A}\equiv\eps(\Gamma^{A})} represents a field-antifield configuration
\mb{\Gamma^{A}\!=\!\left\{\phi^{\alpha};\phi^{*}_{\alpha}\right\}}, 
the antisymplectic structure \mb{E} provides the antibracket
\mb{(\cdot,\cdot)}, and the density \mb{\rho} yields the path integral measure.
However, up until recently, it has been necessary to impose a compatibility
condition \cite{bt93,bbd06} between the two geometric structures \mb{E} and
\mb{\rho} to ensure nilpotency of the odd Laplacian 
\beq
\Deltarho~\equiv~\frac{(-1)^{\eps_{A}}}{2\rho}\lpa{A}\rho E^{AB}\lpa{B}
~,~~~~~~~~~\lpa{A}~\equiv~\papal{\Gamma^{A}}~.\label{deltarho} 
\eeq
In this paper, we show that the compatibility condition between \mb{E} and
\mb{\rho} can be omitted if one adds an odd scalar function \mb{\nu} to the
odd Laplacian \mb{\Deltarho}, 
\beq
\Delta~=~\Deltarho+\nu \label{delta}
\eeq
such that the total \mb{\Delta} operator is nilpotent  
\beq
\Delta^{2}~=~0~.\label{deltanilp}
\eeq
Nilpotency is important for the field-antifield formalism in many ways, for
instance in securing that the physical partition function \mb{{\cal Z}} is 
independent of gauge-choice, see Appendix~\ref{appindep}. (More precisely,
what is really vital is the nilpotency of the underlying \mb{\DeltaE}
operator, \cf Sections~\ref{secdeltae}-\ref{secfindepformalism}.) In physics
terms, the addition of the \mb{\nu} function to the odd Laplacian
\mb{\Deltarho} implies that the quantum master equation
\beq
\Delta e^{\Ih W}~=~0 \label{qme}
\eeq 
is modified with a \mb{\nu} term at the two-loop order
\mb{{\cal O}(\hbar^{2})}:
\beq
\Hf(W,W)~=~i\hbar\Deltarho W+\hbar^{2}\nu~, \label{mqme} 
\eeq
and \mb{\Deltarho} is in general no longer a nilpotent operator.
It turns out that the zeroth-order \mb{\nu} term is uniquely determined
from the nilpotency requirement \e{deltanilp} apart from an odd constant.  
One particular solution to the zeroth-order term, which we call \mb{\nurho}, 
takes a special form \cite{b06}
\beq
\nurho~\equiv~\nu_{\rho}^{(0)}+\frac{\nu^{(1)}}{8}-\frac{\nu^{(2)}}{24}~,
\label{nurho}
\eeq
where \mb{\nu_{\rho}^{(0)}}, \mb{\nu^{(1)}} and \mb{\nu^{(2)}} are defined as
\bea
\nu_{\rho}^{(0)}&\equiv&\frac{1}{\sqrt{\rho}}(\Deltaone\sqrt{\rho})~,
\label{nurho0} \\
\nu^{(1)}&\equiv&
(-1)^{\eps_{A}}(\lpa{B}\lpa{A}E^{AB})~,\label{nu1} \\
\nu^{(2)}&\equiv& (-1)^{\eps_{A}\eps_{C}}(\lpa{D}E^{AB})E_{BC}
(\lpa{A}E^{CD})  \label{nu2} \\
&=&-(-1)^{\eps_{B}}(\lpa{A}E_{BC})
E^{CD}(\lpa{D}E^{BA})~.\label{mnu3} 
\eea
Here, \mb{\Deltaone} in \eq{nurho0} denotes the expression \e{deltarho} for the
odd Laplacian \mb{\Delta_{\rho=1}^{}} with \mb{\rho} replaced by \mb{1}. In
particular, the odd scalar \mb{\nurho} is a function of \mb{E} and \mb{\rho},
so there is no call for new independent geometric structures on the manifold
\mb{M}. In Sections~\ref{secanalysis}--\ref{secnilp2} we show that
\mb{\Deltarho\!+\!\nu} is the only possible \mb{\Delta} operator within the
set of all second-order differential operators. The now obsolete compatibility
condition \cite{bt93,bbd06} between \mb{E} and \mb{\rho} can be recast as
\mb{\nurho={\rm odd~constant}}, thereby making contact to the previous approach
\cite{bt93}, which uses the odd Laplacian \mb{\Deltarho} only. The explicit
formula \e{nurho} for \mb{\nurho} is proven in Section~\ref{secnuf} and
Appendix~\ref{appnuf}. The formula \e{nurho} first appeared in \Ref{b06}. That
paper was devoted to Khudaverdian's \mb{\DeltaE} operator
\cite{k99,kv02,k02,k04}, which takes semidensities to semidensities. This is
no coincidence: At the bare level of mathematical formulas the construction is
intimately related to the \mb{\DeltaE} operator, as shown in
Sections~\ref{secdeltae}-\ref{secfindepformalism}. However the starting point
is different. On one hand, \Ref{b06} studied the \mb{\DeltaE} operator in
its minimal and purest setting, which is a manifold with an antisymplectic
structure \mb{E} but without a density \mb{\rho}. On the other hand,
the starting point of the current paper is a \mb{\Delta} operator that takes
scalar functions to scalar functions, and this implies that a choice of
\mb{\rho} (or \mb{F}, \cf below) should be made.
Later in Sections~\ref{secconn} and \ref{seccurv} we interpret the odd
\mb{\nurho} function as (minus \mb{1/8} times) the odd scalar curvature
\mb{R} of an arbitrary antisymplectic, torsion-free and \mb{\rho}-compatible
connection,
\beq
\nurho~=~-\frac{R}{8}~.\label{rnurho}
\eeq
One of the main priorities for the current article is to ensure that all
arguments are handled in completely general coordinates without resorting to
Darboux coordinates at any stage. This is important to give a physical theory
a natural, coordinate-independent, geometric status in the antisymplectic
phase space. We shall also throughout the paper often address the question of
generalizing the density \mb{\rho} to a non-flat line bundle connection \mb{F}.
It is well-known \cite{bt93} that a density \mb{\rho} gives rise to a flat
line bundle connection
\beq
{}F_{A}~=~(\lpa{A}\ln\rho)~.\label{frhointro}
\eeq
In fact, several mathematical objects, for instance the odd Laplacian
\mb{\Deltarho} and the odd scalar \mb{\nurho}, can be formulated entirely using
\mb{F} instead of \mb{\rho}. Surprisingly, many of these objects continue to be
well-defined for non-flat \mb{F}'s as well, where the nilpotency (and the 
ordinary physical description) is broken down. In Section~\ref{secdnc} we
shall therefore temporarily digress to contemplate a modification of the
nilpotency condition that addresses these mathematical observations. {}Finally,
Section~\ref{secconcl} contains our conclusions.

\noi
{\em General remark about notation}. We have two types of grading: A Grassmann
grading \mb{\eps} and an exterior form degree \mb{p}. The sign conventions
are such that two exterior forms \mb{\xi} and \mb{\eta}, of Grassmann parity
\mb{\eps_{\xi}}, \mb{\eps_{\eta}} and exterior form degree \mb{p_{\xi}},
\mb{p_{\eta}}, respectively, commute in the following graded sense:
\beq
 \eta \wedge \xi
~=~(-1)^{\eps_{\xi}\eps_{\eta}+p_{\xi}p_{\eta}}\xi\wedge\eta
\eeq
inside the exterior algebra.
We will often not write the exterior wedges ``\mb{\wedge}'' explicitly.

\section{General Second-Order \mb{\Delta} operator}
\label{secanalysis}

\noi
We here introduce the setting and notation more carefully, and argue that the
\mb{\Delta} operator must be equal to \mb{\Deltarho\!+\!\nurho} up to an odd
constant. (The undetermined odd constant comes from the fact that the square
\mb{\Delta^{2}=\Hf[\Delta,\Delta]} does not change if \mb{\Delta} is shifted by
an odd constant.) Consider now an arbitrary Grassmann-odd, second-order,
differential operator \mb{\Delta} that takes scalar functions to scalar
functions. In this paper, we shall only discuss the non-degenerate case, where
the second-order term in \mb{\Delta} is of maximal rank, and hence provides for
a non-degenerated antibracket \mb{(\cdot,\cdot)}, \cf the Definition
\e{antibracket} below. (The non-degeneracy assumption is motivated by the fact
that it is satisfied for currently known applications. The degenerate case may
be dealt with via for instance the antisymplectic conversion mechanism
\cite{bm97,b07}.) Due to the non-degeneracy assumption, it is always possible
to organize \mb{\Delta} as
\beq
\Delta~=~\DeltaF+\nu~,\label{deltafnu}
\eeq
where \mb{\nu} is a zeroth-order term and \mb{\DeltaF} is an operator with
terms of second and first order \cite{bt93}
\beq 
\DeltaF~\equiv~\frac{(-1)^{\eps_{A}}}{2}
(\lpa{A}\!+\!F_{A}^{})E^{AB}\lpa{B}~.\label{deltaf}
\eeq
Here, \mb{E^{AB}\!=\!E^{AB}(\Gamma)}, \mb{F_{A}\!=\!F_{A}(\Gamma)} and 
\mb{\nu\!=\!\nu(\Gamma)} is a \mb{(2,0)}-tensor, a line bundle connection, and
a scalar, respectively. We shall sometimes use the slightly longer notation
\mb{\DeltaF\equiv\DeltaFE} to acknowledge that it depends on two inputs: \mb{F}
and \mb{E}. The line bundle connection \mb{F_{A}} transforms under general
coordinate transformations \mb{\Gamma^{A}\to \Gamma^{\prime B}} as 
\beq
{}F_{A}~=~(\papal{\Gamma^{A}}\Gamma^{\prime B}) F^{\prime}_{B}
+(\papal{\Gamma^{A}}\ln J)~,~~~~~~~~
J~\equiv~\sdet \frac{\partial \Gamma^{\prime B}}{\partial\Gamma^{A}}~.
\label{ftransf}
\eeq
These transformation properties guarantee that the expressions
\es{deltafnu}{deltaf} remain invariant under general coordinate
transformations. The Grassmann-parities are
\beq
\eps(E^{AB})~=~\eps_{A}\!+\!\eps_{B}\!+\!1~,~~~~
\eps(F_{A})~=~\eps_{A}~,~~~~\eps(\nu)~=~1~.
\eeq
One may, without loss of generality, assume that the \mb{(2,0)}-tensor
\mb{E^{AB}} has a Grassmann-graded skewsymmetry
\beq
E^{AB}~=~-(-1)^{(\eps_{A}+1)(\eps_{B}+1)}E^{BA}~.\label{skewsym1}
\eeq
The antibracket \mb{(f,g)} of two functions \mb{f=f(\Gamma)} and
\mb{g=g(\Gamma)} is defined via a double commutator\footnote{Here, and
throughout the paper, \mb{[A,B]} and \mb{\{A,B\}} denote the graded commutator
\mb{[A,B]\equiv AB-(-1)^{\eps_{A}\eps_{B}}BA} and the graded anticommutator
\mb{\{A,B\}\equiv AB+(-1)^{\eps_{A}\eps_{B}}BA}, respectively.} \cite{bda96}
with the \mb{\Delta}-operator, acting on the constant unit function \mb{1},
\bea
(f,g)&\equiv&(-1)^{\eps_{f}}[[\stackrel{\rightarrow}{\Delta},f],g]1
~\equiv~(-1)^{\eps_{f}}\Delta(fg) - (-1)^{\eps_{f}}(\Delta f)g
- f(\Delta g) + (-1)^{\eps_{g}}fg (\Delta 1) \cr
&=&(f\rpa{A})E^{AB}(\lpa{B}g)~=~-(-1)^{(\eps_{f}+1)(\eps_{g}+1)}(g,f)~,
\label{antibracket}
\eea
where use is made of the skewsymmetry \e{skewsym1} in the third equality.
By the non-degeneracy assumption, there exists an inverse matrix \mb{E_{AB}} 
such that
\beq
E^{AB}E_{BC}~=~\delta^{A}_{C}~=~E_{CB}E^{BA}~.
\eeq
Since the tensor \mb{E^{AB}} possesses a graded \mb{A\!\leftrightarrow\!B}
skewsymmetry \e{skewsym1}, the inverse tensor \mb{E_{AB}} must be
skewsymmetric,
\beq
E_{AB}=-(-1)^{\eps_{A}\eps_{B}}E_{BA}~. \label{skewsym2}
\eeq
In other words, \mb{E_{AB}} is a two-form
\beq
E~=~\Hf d\Gamma^{A} E_{AB} \wedge d\Gamma^{B}~. \label{etwoform}
\eeq
The Grassmann parity is
\beq
\eps(E_{AB})~=~\eps_{A}\!+\!\eps_{B}\!+\!1~.\label{epslowereab}
\eeq

\section{Nilpotency Conditions: Part I}
\label{secnilp}

\noi
The square \mb{\Delta^{2}=\Hf[\Delta,\Delta]} of an odd second-order operator
\e{deltafnu} is generally a third-order differential operator, which we, for 
simplicity, imagine has been normal ordered, \ie with all derivatives standing
to the right. Nilpotency \e{deltanilp} of the \mb{\Delta} operator leads to
conditions on \mb{E^{AB}}, \mb{F_{A}} and \mb{\nu}. Let us therefore 
systematically, over the next four Sections~\ref{secnilp}--\ref{secnilp2},
discuss order by order the consequences of the nilpotency condition
\mb{\Delta^{2}=0}, starting with the highest (third) order terms, and going
down until we reach the zeroth order. 

\noi
The third-order terms of \mb{\Delta^{2}} vanish if and only if the Jacobi 
identity
\beq
\sum_{{\rm cycl.}~f,g,h}(-1)^{(\eps_{f}+1)(\eps_{h}+1)}
(f,(g,h))~=~0 \label{jacid}
\eeq
for the antibracket \mb{(\cdot,\cdot)} holds. We shall always assume this from
now on. Equivalently, the two-form \mb{E_{AB}} is closed,
\beq
dE~=~0~.\label{eclosed}
\eeq
In terms of the matrices \mb{E^{AB}} and \mb{E_{AB}}, the Jacobi identity
\e{jacid} and the closeness condition \e{eclosed} read 
\bea
\sum_{{\rm cycl.}~A,B,C}(-1)^{(\eps_{A}+1)(\eps_{C}+1)}
E^{AD} (\lpa{D}E^{BC}) &=&0~,\label{abjacid} \\
\sum_{{\rm cycl.}~A,B,C}(-1)^{\eps_{A}\eps_{C}}
(\lpa{A}E_{BC}) &=&0~, \label{abclose}
\eea
respectively. By definition, a non-degenerate tensor \mb{E_{AB}} with 
Grassmann-parity \e{epslowereab}, skewsymmetry \e{skewsym2}, and closeness 
relation \e{abclose} is called an {\em antisymplectic} structure. 

\noi
Granted the Jacobi identity \e{jacid}, the second-order terms of
\mb{\Delta^{2}} can be written on the form
\beq
\frac{1}{4}\cR^{AB}\lpa{B}\lpa{A}~, \label{secondorderterms}
\eeq 
where \mb{\cR^{AB}} with upper indices is a shorthand for
\beq
\cR^{AD}~\equiv~E^{AB}\cR_{BC}E^{CD}(-1)^{\eps_{C}}~,\label{uppercr}
\eeq
and \mb{\cR_{AB}} with lower indices is the curvature tensor for the line
bundle connection \mb{F_{A}}:
\beq
\cR_{AB}^{}~\equiv~[\lpa{A}\!+\!F_{A}^{},\lpa{B}\!+\!F_{B}^{}]
~=~(\lpa{A}F_{B}^{})-(-1)^{\eps_{A}\eps_{B}}(A \leftrightarrow B)~.
\label{crcurvature}
\eeq
Remarkably, the two tensors \mb{\cR_{AB}} and \mb{\cR^{AB}} carry opposite
symmetry:
\bea
\cR_{AB}&=&-(-1)^{\eps_{A}\eps_{B}}\cR_{BA}~,\label{lowercrsymmetry} \\
\cR^{AB}&=&(-1)^{\eps_{A}\eps_{B}}\cR^{BA}~.\label{uppercrsymmetry}
\label{crsymmetry}
\eea
It follows that in the non-degenerate case, the second-order terms of
\mb{\Delta^{2}} vanish if and only if the line bundle connection \mb{F_{A}}
has vanishing curvature
\beq
\cR_{AB}^{}~=~0~.
\label{zerocurvature}
\eeq
The zero curvature condition \e{zerocurvature} is an integrability condition
for the local existence of a density \mb{\rho},
\beq
{}F_{A}~=~(\lpa{A}\ln\rho)~.\label{frho}
\eeq
Under the \mb{F \leftrightarrow\rho} identification \e{frho} the \mb{\DeltaF}
operator \e{deltaf} just becomes the ordinary odd Laplacian \mb{\Deltarho} from
\eq{deltarho},
\beq
\DeltaF~=~\Deltarho~.
\eeq
Conventionally the field-antifield formalism requires the
\mb{F\leftrightarrow\rho} identification \e{frho} to hold globally.
Nevertheless, we shall present many of the constructions below using \mb{F}
rather than \mb{\rho}, to be as general as possible.

\noi
There exists a descriptive characterization: Granted the Jacobi identity
\e{jacid}, the second-order terms of \mb{\Delta^{2}} vanish if and only if
there is a Leibniz rule for the interplay of the so-called ``one-bracket''
\mb{\Phi^{1}_{\Delta}\!\equiv\!\Delta\!-\!(\Delta 1)\!=\!\DeltaF} and the 
``two-bracket'' \mb{(\cdot,\cdot)}
\beq
\DeltaF(f,g)~=~(\DeltaF f,g)-(-1)^{\eps_{f}}(f,\DeltaF g)~.
\label{deltafleibnitz0}
\eeq
See \Ref{bda96,b06cmp} for more details.

\section{A Non-Zero \mb{F}-Curvature?}
\label{secfcurv}

\noi 
In \eq{zerocurvature} of the previous Section~\ref{secnilp} we learned that the
nilpotency condition \e{deltanilp} completely kills the line bundle curvature
\mb{\cR}. Nevertheless, several constructions continue to be well-defined for
non-zero \mb{\cR}. {}For instance, both the important scalars \mb{\nuF} and
\mb{R} fall into this category, \cf \eqs{nuf}{defosc} below. Another example,
which turns out to be related to our discussion, is the Grassmann-odd
\mb{2}-cocycle of Khudaverdian and Voronov \cite{bbd06,kv02,kv04}. It is
defined using two (possibly non-flat) line bundle connections \mb{F^{(1)}} and
\mb{F^{(2)}} as follows:
\beq
\nu(F^{(1)};F^{(2)},E)~\equiv~\frac{1}{4}{\rm div}_{F^{(12)}}^{}X^{}_{(12)}
~\equiv~\frac{(-1)^{\eps_{A}}}{4}
(\lpa{A}+\frac{F^{(1)}+F^{(2)}}{2})(E^{AB}(F_{B}^{(1)}\!-\!F_{B}^{(2)}))~,
\label{nuffprime1}
\eeq
where the divergence ``\mb{{\rm div}}'' is defined in \eq{divf},
\beq
{}F^{(12)}~\equiv~\frac{F^{(1)}+F^{(2)}}{2}~,\label{nuffprime2}
\eeq
and 
\beq
X^{A}_{(12)}~\equiv~E^{AB}(F_{B}^{(1)}\!-\!F_{B}^{(2)})~.\label{nuffprime3}
\eeq
It is clear from Definition \e{nuffprime1} that \mb{\nu(F^{(1)};F^{(2)},E)}
behaves as a scalar under general coordinate transformations. This is because
the average \mb{F^{(12)}} is again a line bundle connection, and \mb{X_{(12)}}
is a vector field since the difference \mb{F_{B}^{(1)}\!-\!F_{B}^{(2)}} is a
co-vector (=one-form), \cf \eq{ftransf}. That \mb{\nu(F^{(1)};F^{(2)},E)} is
a \mb{2}-cocycle
\beq
\nu(F^{(1)};F^{(2)},E)+\nu(F^{(2)};F^{(3)},E)+\nu(F^{(3)};F^{(1)},E)~=~0
\label{twococycle}
\eeq
follows easily by rewriting Definition \e{nuffprime1} as
\beq
\nu(F^{(1)};F^{(2)},E)~=~\nu_{F^{(1)}}^{(0)}-\nu_{F^{(2)}}^{(0)}~,
\label{twocoboundary}
\eeq
where \mb{\nu_{F}^{(0)}} generalizes \eq{nurho0}:
\beq
\nu_{F}^{(0)}~\equiv~
\frac{(-1)^{\eps_{A}}}{4}(\lpa{A}+\frac{F_{A}}{2})(E^{AB}F_{B})~.
\label{nuf0}
\eeq
Note that Definitions \es{nuffprime1}{nuf0} continue to make
sense for non-flat \mb{F}'s. We should stress that \mb{\mb{\nu_{F}^{(0)}}}
itself is {\em not} a scalar, but we shall soon see that it can be replaced in
\eq{twocoboundary} by a scalar \mb{\nuF}, \cf \eq{nuf} below. In other words, 
\mb{\nu(F^{(1)};F^{(2)},E)} is a \mb{2}-coboundary.

\noi
The \mb{F}-curvature \mb{\cR_{AB}} is also an interesting geometric object in
its own right. It can be identified with a Ricci two-form of a tangent bundle
connection \mb{\nabla}, \cf \eq{ricciform} in Section~\ref{seccurv} below.
The Ricci two-form
\beq 
\cR~=~\Hf d\Gamma^{A}\cR_{AB}^{}\wedge d\Gamma^{B}(-1)^{\eps_{B}}
\label{crtwoform}
\eeq
is closed 
\beq
d\cR~=~0~,\label{crclosed}
\eeq
due to the Bianchi identity
\beq
\sum_{{\rm cycl.}~A,B,C}(-1)^{\eps_{A}\eps_{C}}
(\lpa{A}\cR_{BC}^{})~=~0~, \label{crbianchi}
\eeq
so the two-form \e{crtwoform} defines a cohomology class.

\section{Breaking the Nilpotency}
\label{secdnc}

\noi
Due to the above mathematical reasons we shall digress in this
Section~\ref{secdnc} to contemplate how a non-zero \mb{F}-curvature could
arise in antisymplectic geometry, although we should stress that it remains
unclear if it is useful in physics. Nevertheless, the strategy that we shall
adapt here is to append a general Grassmann-even (possibly degenerate) 
second-order operator source \mb{\Hf\DeltaR} to the \rhs of the nilpotency
condition \e{deltanilp}:
\beq
\Delta^{2}~=~\Hf\DeltaR~.\label{newdeltasquarecond}
\eeq
A covariant and general way of realizing the second-order \mb{\DeltaR}
operator is to write
\beq
\DeltaR~\equiv~\DeltaFR + \VR + \nR~,
\label{deltar}
\eeq
where 
\beq
\DeltaFR~\equiv~\frac{(-1)^{\eps_{A}}}{2}
(\lpa{A}\!+\!F_{A}^{})\cR^{AB}\lpa{B} \label{deltafr}
\eeq
is an Grassmann-even Laplacian based on \mb{F_{A}} and \mb{\cR^{AB}}. We have
included a Grassmann-even vector field
\beq
\VR~\equiv~V_{\cR}^{A}\lpa{A}
\eeq
and a scalar function \mb{\nR} to give a systematic treatment. Note that the
vector field \mb{\VR} is the difference of the subleading connection terms
inside \mb{\DeltaR} and \mb{\DeltaFR}. We shall show below that the
\mb{\nR} term is completely determined by consistency, while \mb{\VR} in
principle can be any locally Hamiltonian vector field subjected to the
following restriction: Both \mb{V_{\cR}^{A}} and \mb{\nR} should be
proportional to the \mb{\cR}-source (or its derivatives) in order to restore
nilpotency \e{deltanilp} in the limit \mb{\cR \to 0}. 

\noi
The new condition \e{newdeltasquarecond} still imposes the Jacobi identity
\e{jacid} for the antibracket \mb{(\cdot,\cdot)} at the third order, since the
modification is just of second order. (We mention, for later, that the Jacobi
identity alone guarantees the existence of a nilpotent \mb{\DeltaE} operator
and its quantization scheme, \cf
Sections~\ref{secdeltae}-\ref{secfindepformalism}, regardless of how the
nilpotency \e{newdeltasquarecond} of \mb{\Delta} is broken at lower orders.) 
The second-order terms in \eq{newdeltasquarecond} implies that the 
\mb{F}-curvature \mb{\cR^{AB}} defined in \eq{crcurvature} should be identified
with the principal symbol \mb{\cR^{AB}} appearing inside the \mb{\DeltaFR}
operator \e{deltafr}, thereby justifying the notation. Note that the Leibniz
rule \e{deltafleibnitz0} is {\em no} longer valid. To see this, it is useful to
define an even \mb{\cR}-bracket \cite{b97}
\bea
(f,g)_{\cR}^{}&\equiv&[[\stackrel{\rightarrow}{\Delta}_{\cR},f],g]1 
~\equiv~\DeltaR(fg) - (\DeltaR f)g - f(\DeltaR g) + fg (\DeltaR 1) \cr
&=&(f\rpa{A})\cR^{AB}(\lpa{B}g)
~=~(-1)^{\eps_{f}\eps_{g}}(g,f)_{\cR}^{}~.\label{rbracket}
\eea
It turns out that the \mb{\cR}-bracket \mb{(\cdot,\cdot)_{\cR}^{}} measures the
failure of the Leibniz rule:
\beq
\Hf(f,g)_{\cR}^{}~=~(-1)^{\eps_{f}}\DeltaF(f,g)-(-1)^{\eps_{f}}(\DeltaF f,g)
+(f,\DeltaF g)~.
\label{deltafleibnitz1}
\eeq
Note that this \mb{\cR}-bracket \mb{(\cdot,\cdot)_{\cR}^{}} does {\em not}
satisfy a Jacobi identity. (In fact, we shall see that the closeness relation
\e{crclosed} for \mb{\cR_{AB}} will instead lead to a compatibility relation
\e{frrel1} below.) Since \mb{\Delta_{F}^{2}\!-\!\Hf\DeltaFR} is a first-order
operator, \cf \eqs{deltafnu}{newdeltasquarecond}, the commutator
\beq
\Hf [\DeltaFR,\DeltaF] ~=~ [\DeltaF,\Delta_{F}^{2}\!-\!\Hf\DeltaFR]
\label{prelemmaa}
\eeq 
becomes a second-order operator at most. (We shall improve this estimate in
Lemma~\ref{lemmaa} below.) This fact already implies that the two brackets
\mb{(\cdot,\cdot)} and \mb{(\cdot,\cdot)_{\cR}^{}} are compatible in the sense
that
\beq
\sum_{{\rm cycl.}~f,g,h} (-1)^{\eps_{f}(\eps_{h}+1)} ((f,g),h)_{\cR}^{}
~=~ \sum_{{\rm cycl.}~f,g,h}(-1)^{\eps_{f}(\eps_{h}+1)+\eps_{g}} 
((f,g)_{\cR}^{},h)~.\label{frrel1}
\eeq
Phrased differently, one may define a one-parameter family of antisymplectic
two-forms 
\beq
E(\theta)~\equiv~E+\theta\cR~\equiv~E+\cR\theta
~=~\Hf d\Gamma^{A} E_{AB}(\theta) \wedge d\Gamma^{B}
~,~~~~~~~~~~~~~~~dE(\theta)~=~0~,
\eeq 
which depends on a Grassmann-odd parameter \mb{\theta}. In components it reads
\bea
E_{AB}(\theta)&=&E_{AB}+\cR_{AB}\theta~, \\
E^{AB}(\theta)&=&E^{AB}+(-1)^{\eps_{A}}\theta\cR^{AB}
~=~E^{AB}+\cR^{AB}\theta(-1)^{\eps_{B}}~.
\eea
There exists locally an antisymplectic one-form potential
\beq
\begin{array}{rclcrcl}
U(\theta)&\equiv&U_{A}^{}(\theta) d\Gamma^{A}~,&~~~~~&
U_{A}^{}(\theta)&\equiv&U_{A}^{}+F_{A}^{}\theta~,
 \cr\cr
dU(\theta)&=&E(\theta)~, &&
\lpa{A}U_{B}^{}(\theta)-(-1)^{\eps_{A}\eps_{B}}(A \leftrightarrow B)
&=&E_{AB}^{}(\theta)~.
\end{array}
\eeq
We will now improve the estimate from \eq{prelemmaa}:

\begin{lemma}
The commutator \mb{[\DeltaF,\DeltaFR]} is always a first-order operator at
most.
\label{lemmaa}
\end{lemma}

\noi
{\sc Proof of Lemma~\ref{lemmaa}:}~~Note that the commutator
\mb{[\DeltaF,\DeltaFR]} appears inside the square 
\beq
\left(\Delta_{F}(\theta)\right)^{2}
~=~\Delta_{F}^{2}+\theta[\DeltaFR,\DeltaF] 
~=~\Delta_{F}^{2}+[\DeltaF,\DeltaFR]\theta
\label{squaretheta}
\eeq
of the Grassmann-odd second-order operator
\beq
\DeltaF(\theta)~\equiv~\DeltaF+\theta\DeltaFR~\equiv~\DeltaF+\DeltaFR\theta
~=~\frac{(-1)^{\eps_{A}}}{2}(\lpa{A}\!+\!F_{A}^{})E^{AB}(\theta)\lpa{B}~.
\eeq
One knows from the general discussion in the previous Section~\ref{secnilp}
that the third-order terms in the square \e{squaretheta} vanish because
\mb{E^{AB}(\theta)} satisfies the Jacobi identity \e{abjacid}. Moreover, the
second-order terms in the square \e{squaretheta} are of the form
\beq
\frac{(-1)^{\eps_{C}}}{4}E^{AB}(\theta)~
\cR_{BC}~E^{CD}(\theta)~\lpa{D}\lpa{A}
~=~\frac{1}{4}\cR^{AB}\lpa{B}\lpa{A}~, \label{secondordertermstheta}
\eeq 
\cf \eqs{secondorderterms}{uppercr}. It is easy to see that the two
\mb{\theta}-dependent terms inside the \lhs of \eq{secondordertermstheta}
cancel against each other. In fact, each of the two terms vanish separately
due to skewsymmetry:
\beq
(-1)^{\eps_{C}+\eps_{F}}
E^{AB}\cR_{BC}E^{CD}\cR_{DF}E^{FG}~=~\cR^{AC}E_{CD}\cR^{DG}
~=~(-1)^{(\eps_{A}+1)(\eps_{G}+1)}(A\leftrightarrow G)~.
\eeq
Therefore, the \mb{\theta}-dependent part of the square \e{squaretheta} must
be of first order at most.
\proofbox

\noi
(One may also give a proof of Lemma~\ref{lemmaa} based on Lemma~\ref{lemmab} in
Appendix~\ref{appnuf}.) Lemma~\ref{lemmaa} implies (for instance via the
technology of \Ref{bda96}) that
\bea
\DeltaFR(f,g) -(\DeltaFR f,g)-(f,\DeltaFR g)&=&
(-1)^{\eps_{f}}\DeltaF(f,g)_{\cR}^{} 
-(-1)^{\eps_{f}}(\DeltaF f,g)_{\cR}^{}\cr 
&&-(f,\DeltaF g)_{\cR}^{}~,\label{frrel2} \\
(\Delta_{F}^{2}\!-\!\Hf\DeltaFR)(f,g)
&=&((\Delta_{F}^{2}\!-\!\Hf\DeltaFR)f,g)
+(f,(\Delta_{F}^{2}\!-\!\Hf\DeltaFR)g)~.\label{frrel3}
\eea
More generally, there exists a superformulation
\beq
\Delta(\theta)~\equiv~\Delta+\theta\DeltaR~\equiv~\Delta+\DeltaR\theta
~=~\frac{(-1)^{\eps_{A}}}{2}(\lpa{A}\!+\!F_{A}^{}(\theta))
E^{AB}(\theta)\lpa{B}+\nu(\theta)~ \label{deltatheta}~,
\eeq
where
\beq
\nu(\theta)~\equiv~\nu+\theta\nR~\equiv~\nu+\nR\theta~,\label{nutheta}
\eeq
and
\beq
{}F_{A}^{}(\theta)~\equiv~F_{A}^{}+2E_{AB}V_{\cR}^{B}\theta
~\equiv~F_{A}^{}-2V_{\cR}^{B}E_{BA}\theta~. \label{ftheta}
\eeq
The nilpotency condition
\beq
\left(\Delta(\theta)-\Hf\frac{\partial }{\partial\theta}\right)^{2}~=~0~
\eeq
precisely encodes the deformed condition \e{newdeltasquarecond} and its 
consistency relation
\bea
0&=&[\Delta,[\Delta,\Delta]]~=~[\Delta,\DeltaR]~=~
[\DeltaF\!+\!\nu,\DeltaFR\!+\!\VR\!+\!\nR] \cr
&=&[\DeltaF,\DeltaFR]+[\DeltaF,\VR]
+[\DeltaF,\nR]-[\DeltaFR\!+\!\VR,\nu]~.\label{consisrel}
\eea
Note in the last line of \eq{consisrel} that the first term
\mb{[\DeltaF,\DeltaFR]} and the two last terms \mb{[\DeltaF,\nR]} and
\mb{[\DeltaFR\!+\!\VR,\nu]} are all of first order. Hence, the second term
\mb{[\DeltaF,\VR]} must be of first order as well. This in turn implies that
\mb{\VR} should be a generating vector field for an anticanonical
transformation:
\beq
\VR(f,g)~=~(\VR(f),g)+(f,\VR(g))~. 
\eeq
Since the antibracket is non-degenerated, it follows that \mb{\VR} must be a
locally Hamiltonian vector field, which we, for simplicity, will assume is a
globally Hamiltonian vector field 
\beq 
\VR~=~-2(\nuR,\cdot)~,   \label{hvf}
\eeq
with some Fermionic globally defined Hamiltonian \mb{\nuR}. The factor
``\mb{-2}'' in \eq{hvf} is chosen for later convenience. The Hamiltonian
\mb{\nuR} in \eq{hvf} should be considered as an additional geometric input,
which labels the different ways \e{newdeltasquarecond} of breaking the
nilpotency of \mb{\Delta}. It is a priori only defined in \eq{hvf} up to an
odd constant. We fix this constant by requiring that 
\beq
\nuR~\to~0~~~~~~ {\rm for}~~~~~~\cR~\to~0~. \label{nurbc}
\eeq
Altogether, the Hamiltonian  \mb{\nuR} does not contribute to the
curvature
\beq
\lpa{A}F_{B}^{}(\theta)
-(-1)^{\eps_{A}\eps_{B}}(A \leftrightarrow B)
~=~\cR_{AB}^{}
\eeq
of the line bundle connection
\beq
{}F_{A}^{}(\theta)~=~F_{A}^{}+4(\lpa{A}\nuR)\theta~.
\eeq
Now let us continue the investigation of the deformed condition
\e{newdeltasquarecond}. The first-order terms of \eq{newdeltasquarecond} cancel
if and only if
\beq
\Delta_{F}^{2}-\Hf\DeltaFR~=~(\nu\!-\!\nuR,\cdot)~. \label{nueq}
\eeq
This is a differential equation for the function \mb{\nu\!=\!\nu(\Gamma)}, or,
equivalently, for the difference \mb{\nu\!-\!\nuR}.
It now becomes clear that the \mb{\nuR} function provides an auxiliary
curvature background for the \mb{\nu} function. Since we assume that \mb{\nuR}
is given, we will now focus on the difference \mb{\nu\!-\!\nuR} rather than on
\mb{\nu} itself. The Frobenius integrability condition for \eq{nueq} comes
from the fact that the operator \mb{\Delta_{F}^{2}-\Hf\DeltaFR}
differentiates the antibracket, \cf \eq{frrel3}. This implies that the
difference \mb{\nu\!-\!\nuR} can be written as a contour integral
\beq
 (\nu\!-\!\nuR)(\Gamma)
~=~(\nu\!-\!\nuR)(\Gamma_{0})+\int_{\Gamma_{0}}^{\Gamma} \!\!
\left. ((\Delta_{F}^{2}\!-\!\Hf\DeltaFR)\Gamma^{A})E_{AB}
\right|_{\Gamma\to\Gamma^{\prime}}d\Gamma^{\prime B}
\label{integralformula}
\eeq
that is independent of the curve (aside from the two endpoints). It only
depends on \mb{E}, \mb{F}, and an odd integration constant
\mb{(\nu\!-\!\nuR)(\Gamma_{0})}. In particular, we conclude that the difference
\mb{\nu\!-\!\nuR} does not introduce any new geometric structures. The
first-order commutator from Lemma~\ref{lemmaa} can now be expressed in terms
of the difference \mb{\nu\!-\!\nuR} as follows:
\bea
\Hf[\DeltaFR,\DeltaF]&=&[\DeltaF,\Delta_{F}^{2}-\Hf\DeltaFR] 
~=~\DeltaF(\nu\!-\!\nuR,\cdot)-(\nu\!-\!\nuR,\DeltaF(\cdot)) \cr
&=&(\DeltaF(\nu\!-\!\nuR),\cdot)-\Hf (\nu\!-\!\nuR,\cdot)_{\cR}^{}~.
\label{commutatorformel}
\eea
Here, \eq{nueq} is used in the second equality and the deformed Leibniz rule
\e{deltafleibnitz1} is used in the third (=last) equality.
 
\noi
{}Finally, the zeroth-order terms of \eq{newdeltasquarecond} cancel if and only
if
\beq
\nR~=~2(\DeltaF\nu)~, \label{zerothorder}
\eeq
so this fixes completely the Grassmann-even function \mb{\nR}. 
One can show that if the Hamiltonian vector field \mb{V_{\cR}^{A}} vanishes
in the flat limit \mb{\cR\to0}, then the \mb{\nR} function, defined via
\eq{zerothorder}, automatically does the same, \cf \eq{zerothorder0proof}
below. The nilpotency-breaking operator \mb{\DeltaR} will therefore vanish for
\mb{\cR\to0}, as it should.

\section{Nilpotency Conditions: Part II}
\label{secnilp2}

\noi
After this digression into non-zero \mb{\cR} curvature, let us now return to
the nilpotent (and ordinary physical) situation \mb{\Delta^{2}=0}, where
\mb{\cR}, \mb{V_{\cR}^{A}} and \mb{\nR} are all zero. Not much changes for the
condition \e{nueq} for the first-order terms other that one should remove the
\mb{\nuR} function and the \mb{\DeltaFR} operator from the Frobenius
integrability condition \e{frrel3}, the differential \eq{nueq}, and the contour
integral \e{integralformula}. (Of course, now the Frobenius integrability
condition is just an easy consequence of the Leibniz rule \e{deltafleibnitz0}
applied twice.) The condition \e{zerothorder} for the zeroth-order terms
becomes
\beq
(\DeltaF\nu)~=~0~. \label{zerothorder0}
\eeq
Equation \e{zerothorder0} is not an independent condition but it follows
instead automatically from the previous requirements.~~{\sc Proof}:
\bea
-(\DeltaF\nu)
&=&\frac{(-1)^{\eps_{A}}}{2}(\lpa{A}\!+\!F_{A}^{})(\nu,\Gamma^{A}) 
~=~\frac{(-1)^{\eps_{A}}}{2}
(\lpa{A}\!+\!F_{A}^{})\Delta_{F}^{2}\Gamma^{A} \cr 
&=&\frac{(-1)^{\eps_{A}+\eps_{B}}}{4}
(\lpa{A}\!+\!F_{A}^{})(\lpa{B}\!+\!F_{B}^{})
(\Gamma^{B},\DeltaF\Gamma^{A}) \cr
&=&-\frac{(-1)^{\eps_{A}}}{8}(\lpa{A}\!+\!F_{A}^{})(\lpa{B}\!+\!F_{B}^{})
\DeltaF(\Gamma^{B},\Gamma^{A}) \cr
&=&\frac{(-1)^{\eps_{A}\eps_{C}}}{16}
(\lpa{A}\!+\!F_{A}^{})(\lpa{B}\!+\!F_{B}^{})(\lpa{C}\!+\!F_{C}^{})
(\Gamma^{C},(\Gamma^{B},\Gamma^{A}))
(-1)^{(\eps_{A}+1)(\eps_{C}+1)}~=~0~. \label{zerothorder0proof}
\eea
Here, the \mb{\nu} \eq{nueq} is used in the second equality, the Leibniz rule
\e{deltafleibnitz0} in the fourth equality, the Jacobi identity \e{jacid} in
the sixth (=last) equality, and the zero curvature condition \e{zerocurvature}
in the second, fourth and sixth equality.
\proofbox

\section{An Explicit Solution \mb{\nuF}}
\label{secnuf}

\noi
Remarkably, the integral \e{integralformula} can be performed. 

\begin{proposition}
The odd quantity
\beq
\nuF~\equiv~\nu_{F}^{(0)}+\frac{\nu^{(1)}}{8}-\frac{\nu^{(2)}}{24}
\label{nuf} 
\eeq
is a solution to the differential \eq{nueq} for the difference
\mb{\nu\!-\!\nuR}, even if the line bundle connection \mb{F} is not flat.
\label{propositiona}
\end{proposition}

\noi
Here, \mb{\nu_{F}^{(0)}}, \mb{\nu^{(1)}} and \mb{\nu^{(2)}} are given by eqs.\
\e{nuf0}, \es{nu1}{nu2}, respectively. Proposition~\ref{propositiona} is proven
in Appendix~\ref{appnuf} by repeated use of the Jacobi identity \e{abjacid} and
the closeness relation \e{abclose}. Notice that under the
\mb{F\leftrightarrow\rho} identification \e{frho}, the \mb{F}-dependent
Definitions \es{nuf0}{nuf} reduce to their \mb{\rho} counterparts
\es{nurho}{nurho0},
\beq
\nuF~=~\nurho~,~~~~~~~~\nu_{F}^{(0)}~=~\nu_{\rho}^{(0)}~.
\eeq
{\em Notation}: \mb{\nuF} or \mb{\nurho} with subscript ``\mb{F}'' or 
``\mb{\rho}'' denotes one particular solution \e{nuf} or \e{nurho} to the
difference \mb{\nu\!-\!\nuR} in \eq{nueq}, respectively.
 
\begin{proposition}
The \mb{\nuF} quantity \e{nuf} is invariant under general coordinate
transformations, \ie it is a scalar, even if the line bundle connection \mb{F}
is not flat.
\label{propositionb}
\end{proposition}

\noi
{\sc Proof of Proposition~\ref{propositionb}}:~~Under an arbitrary
infinitesimal coordinate transformation \mb{\delta\Gamma^{A}=X^{A}}, one
calculates \cite{b06}
\bea
\delta\nu_{F}^{(0)}&=&-\Hf \Deltaone{\rm div}_{1}^{}X~, \label{dnuf0}  \\
\delta\nu^{(1)}&=& 4 \Deltaone{\rm div}_{1}^{}X 
+ (-1)^{\eps_{A}}(\lpa{C}E^{AB})
(\lpa{B}\lpa{A}X^{C})~, \label{dnu1} \\
\delta\nu^{(2)}&=&3(-1)^{\eps_{A}}(\lpa{C}E^{AB})(\lpa{B}\lpa{A}X^{C})~,
\label{dnu2}
\eea
where \mb{\Deltaone} and \mb{{\rm div}_{1}^{}} denote the expressions
\es{deltarho}{divrho} for the odd Laplacian \mb{\Delta_{\rho=1}^{}} and the
divergence \mb{{\rm div}_{\rho=1}} with \mb{\rho} replaced by \mb{1}.
One easily sees that while the three constituents \mb{\nu_{F}^{(0)}},
\mb{\nu^{(1)}} and \mb{\nu^{(2)}} separately have non-trivial transformation
properties, the linear combination \mb{\nuF} in \eq{nuf} is indeed a scalar.
Proposition~\ref{propositionb} also follows from the identification of
\mb{\nuF} as an odd scalar curvature, \cf \eq{rnuf} below. 
\proofbox

\noi
The difference \mb{\nu\!-\!\nuR} is only determined up to an odd integration
constant because the defining relation \e{nueq} is a differential relation. The
explicit solution \mb{\nuF} in \e{nuf} provides us with an opportunity to fix
this odd integration constant once and for all. Out of all the solutions to
the difference \mb{\nu\!-\!\nuR}, we choose the \mb{\nuF} solution \e{nuf},
\ie we identify from now on 
\beq
\nu~\equiv~\nuF+\nuR~.
\eeq
We do this for two reasons.
{}Firstly, any odd
constants inside the \mb{\nuF} expression \e{nuf} can only arise implicitly
through \mb{E} and \mb{F}, which means that if \mb{E} and \mb{F} do not carry
any odd constants, then the \mb{\nuF} solution \e{nuf} will be free of odd
constants as well. Similarly, the \mb{\nuR} part does not contain odd constants
because of the boundary condition \e{nurbc}. Secondly, the expression \mb{\nuF}
is the only solution that has an interpretation as an odd scalar curvature,
\cf \eq{rnuf} below. This completes the reduction of a general second-order
\mb{\Delta} operator to
\beq
\Delta~=~\DeltaF+\nu
~=~\DeltaF+\nuF+\nuR ~~~\longrightarrow~~~ \Deltarho+\nurho~~~~~~~
{\rm for}~~~~~~~\cR ~\to~ 0~. 
\eeq

\section{The \mb{\DeltaE} operator}
\label{secdeltae}

\noi
Let us briefly outline the connection to Khudaverdian's \mb{\DeltaE} operator
\cite{k99,kv02,k02,k04}, which takes semidensities to semidensities.
The \mb{\DeltaE} operator was defined in \Ref{b06} as
\beq
\DeltaE~\equiv~\Deltaone+\frac{\nu^{(1)}}{8}-\frac{\nu^{(2)}}{24}~, 
\label{deltaedef}
\eeq
where \mb{\Deltaone} denotes the expression \e{deltarho} for the odd Laplacian
\mb{\Delta_{\rho=1}^{}} with \mb{\rho} replaced by \mb{1}.
Some of the strengths of Definition \e{deltaedef} are that it works in any
coordinate system and that it is manifestly independent of \mb{\rho} or \mb{F}.
However, it is a rather lengthy calculation to demonstrate in a \mb{\rho}-less
or \mb{F}-less environment that \mb{\DeltaE} has the pertinent transformation
property under general coordinate transformations, and that it is nilpotent
\beq
\Delta_{E}^{2}~=~0~, \label{deltaenilp}
\eeq
\cf \Ref{b06}. Once we are given a density \mb{\rho}, the situation simplifies
considerably. Then, the \mb{\DeltaE} operator becomes just the operator
\mb{\Delta\equiv\Deltarho\!+\!\nurho} conjugated with the square root of
\mb{\rho}:
\beq
\DeltaE~=~\sqrt{\rho}\Delta\frac{1}{\sqrt{\rho}}~. \label{deltaedeltarho}
\eeq
{\sc Proof of \eq{deltaedeltarho}}:~~Let \mb{\sigma} denote an arbitrary
semidensity. Then, it follows from the explicit \mb{\nurho} formula \e{nurho}
that
\bea
(\DeltaE\sigma)
&=&(\Deltaone\sigma)+(\frac{\nu^{(1)}}{8}-\frac{\nu^{(2)}}{24})\sigma
~=~(\Deltaone\sigma)-(\Deltaone\sqrt{\rho})\frac{\sigma}{\sqrt{\rho}}
+\nurho\sigma \cr
&=&\sqrt{\rho}(\Deltaone\frac{\sigma}{\sqrt{\rho}}) 
+(\sqrt{\rho},\frac{\sigma}{\sqrt{\rho}})+\nurho\sigma
~=~\sqrt{\rho}(\Deltarho\frac{\sigma}{\sqrt{\rho}})+\nurho\sigma 
~=~\sqrt{\rho}(\Delta\frac{\sigma}{\sqrt{\rho}})~.
\eea
It is remarkable that the \mb{\sqrt{\rho}}-conjugated \mb{\Delta} operator
\mb{\sqrt{\rho}\Delta\frac{1}{\sqrt{\rho}}} does not depend on \mb{\rho} at
all! On the other hand, it is obvious that the operator
\mb{\sqrt{\rho}\Delta\frac{1}{\sqrt{\rho}}} is nilpotent and that it satisfies
the required transformation law under general coordinate transformations, \ie
that it takes semidensities to semidensities. This is because the \mb{\Delta}
operator itself is a nilpotent operator and \mb{\Delta} takes scalar functions
to scalar functions. Let us also mention that
\beq
\nurho~=~(\Delta 1)~=~\frac{1}{\sqrt{\rho}}(\DeltaE\sqrt{\rho})~.
\label{nurhofromdeltae}
\eeq
The \rhs of \eq{nurhofromdeltae} served as a definition of the odd scalar
\mb{\nurho} in \Ref{b06}.

\noi
More generally, the operators \mb{\DeltaE} and
\mb{\Delta\equiv\DeltaF\!+\!\nuF\!+\!\nuR} are linked via
\beq
\DeltaE~=~\Delta-\frac{(-1)^{\eps_{A}}}{2}F_{A}(\Gamma^{A},\cdot)
-\nu_{F}^{(0)}-\nuR~. \label{deltaedeltaf}
\eeq 
Equation \e{deltaedeltaf} may be viewed as a generalization of
\eq{deltaedeltarho} to non-flat \mb{F}'s, or, equivalently, to non-nilpotent
\mb{\Delta}'s, \cf \eq{zerocurvature}. It might be worth emphasizing that
\mb{\DeltaE} is nilpotent even in this situation, since \mb{\DeltaE} only
depends on \mb{E}.

\section{\mb{F}-Independent Formalism}
\label{secfindepformalism}

\noi
There exists \cite{b06,b07} a manifestly \mb{F}-independent quantization scheme
based on the \mb{\DeltaE} operator. Since we will demand that the quantization
is covariant \wrt the antisymplectic phase space, it will be necessary to use 
first-level formalism or one of its higher-level generalizations
\cite{bt93,bt94}. See \Ref{bbd06} for a review of the multi-level formalism.
It turns out to be most efficient to use the second-level formalism in order
not to deal directly with weak quantum master equations \cite{bms95}. Let
\mb{\Gamma^{A}} denote all the zeroth- and first-level fields and antifields,
and let \mb{\lambda^{\alpha}} denote the second-level Lagrange multipliers for
the first-level gauge-fixing constraints. Assume also that there is no
dependence on the corresponding second-level antifields
\mb{\lambda^{*}_{\alpha}}. The second-level partition function
\beq
{\cal Z}~=~\int\! [d\Gamma][d\lambda]~e^{\Ih(\WE+\XE)}
\label{epartitionfct}
\eeq 
contains two Boltzmann semidensities: a gauge-generating semidensity
\mb{e^{\Ih\WE}} and a gauge-fixing semidensity \mb{e^{\Ih\XE}}, where \mb{\WE}
and \mb{\XE} denote the corresponding quantum actions. The two Boltzmann
semidensities are both required to satisfy strong quantum master equations
\beq
\DeltaE e^{\Ih\WE}~=~0~,~~~~~~~~~~~~\DeltaE e^{\Ih\XE}~=~0~,
\label{eqmewx}
\eeq
or equivalently,
\beq
\Hf(\WE,\WE)~=~i\hbar\Deltaone\WE+\hbar^{2}\DeltaE 1~,~~~~~~~~~
\Hf(\XE,\XE)~=~i\hbar\Deltaone\XE+\hbar^{2}\DeltaE 1~,
\label{emqmewx} 
\eeq 
where
\beq
\DeltaE 1~=~\frac{\nu^{(1)}}{8}-\frac{\nu^{(2)}}{24}~.\label{deltaeone} 
\eeq
The caveat is that the quantum actions \mb{\WE} and \mb{\XE} are {\em not}
scalars. They obey non-trivial transformation laws under general coordinate
transformations, since they are logarithms of semidensities.  It is shown in
Appendix~\ref{appindep} that the partition function \e{epartitionfct} is
independent of the gauge choice \mb{\XE}. 

\noi
If we are given a density \mb{\rho}, we may introduce a nilpotent \mb{\Delta}
operator \e{deltaedeltarho} and Boltzmann scalars \mb{e^{\Ih W}} and
\mb{e^{\Ih X}} by dressing appropriately with square roots of \mb{\rho}:
\beq
\sqrt{\rho}\Delta~=~\DeltaE \sqrt{\rho}~,~~~~~~~~~~  
e^{\Ih\WE}~=~\sqrt{\rho}e^{\Ih W}~,~~~~~~~~~~
e^{\Ih\XE}~=~\sqrt{\rho}e^{\Ih X}~.\label{eidentifications}
\eeq
Then \mb{\Delta=\Deltarho\!+\!\nurho} and the two scalar actions \mb{W} and
\mb{X} will satisfy the strong quantum master \eq{qme} from the Introduction,
which in non-exponential form reads
\beq
\Hf(W,W)~=~i\hbar\Deltarho W+\hbar^{2}\nurho~,~~~~~~~~~~~~
\Hf(X,X)~=~i\hbar\Deltarho X+\hbar^{2}\nurho~. \label{mqmewx} 
\eeq 
The partition function \e{epartitionfct} then reduces to the familiar
\mb{W}-\mb{X} form:
\beq
{\cal Z}~=~\int\! \rho [d\Gamma][d\lambda]~e^{\Ih (W+X)}~.
\label{rhopartitionfct}
\eeq
Conversely, since the partition function \e{rhopartitionfct} via the above 
identifications \e{eidentifications} can be written in the manifestly
\mb{\rho}-independent form \e{epartitionfct}, one may state that in this sense
the partition function \e{rhopartitionfct} does not depend on \mb{\rho}. The
point is that the well-known ambiguity in the choice of measure that exists in
the field-antifield formalism has been fully transcribed into an ambiguity in
the choice of the Boltzmann semidensity \mb{e^{\Ih\WE}}. Put differently, if
one splits the Boltzmann semidensity \mb{e^{\Ih\WE}} into a Boltzmann scalar
\mb{e^{\Ih W}} and a density \mb{\rho} as done in \eq{eidentifications}, the 
measure ambiguity sits inside the scalar \mb{e^{\Ih W}}, not in \mb{\rho},
as \mb{\rho} actually drops out of \mb{{\cal Z}}.

\noi   
More generally, imagine that we are given a non-nilpotent operator
\mb{\Delta\equiv\DeltaF\!+\!\nuF\!+\!\nuR} with a non-flat line bundle
connection \mb{F} that satisfies the deformed nilpotency condition
\e{newdeltasquarecond}. We can still define the partition function in this
situation via the above quantization scheme \e{epartitionfct} based on the
nilpotent \mb{\DeltaE} operator. Such an approach will of course be manifestly
\mb{F}-independent by construction.

\section{Connection}
\label{secconn}

\noi
We now introduce a connection \mb{\nabla:TM\times TM\to TM}. See
\Ref{b97,lavrov04} for related discussions. The left covariant derivative
\mb{(\nabla_{A}X)^{B}} of a left vector field \mb{X^{A}} is defined as
\cite{b97}
\beq
(\nabla_{A}X)^{B}~\equiv~(\lpa{A}X^{B})
+(-1)^{\eps_{X}(\eps_{B}+\eps_{C})}\Gamma_{A}{}^{B}{}_{C}X^{C}~,~~~~~~~
\eps(X^{A})~=~\eps_{X}+\eps_{A}~,\label{nabladef}
\eeq 
The word ``left'' implies that \mb{X^{A}} and \mb{(\nabla_{A}X)^{B}} transform
with left derivatives
\beq
X^{\prime B}~=~X^{A}(\papal{\Gamma^{A}}\Gamma^{\prime B})~,~~~~~~~~~
(\papal{\Gamma^{A}} \Gamma^{\prime B})(\nabla_{\prime B}X)^{\prime C}
~=~(\nabla_{A}X)^{B}(\papal{\Gamma^{B}}\Gamma^{\prime C})~,
\eeq
under general coordinate transformations \mb{\Gamma^{A}\to \Gamma^{\prime B}}.
It is convenient to introduce a reordered Christoffel symbol
\beq
\Gamma^{A}{}_{BC}~\equiv~(-1)^{\eps_{A}\eps_{B}}\Gamma_{B}{}^{A}{}_{C}
\label{alternativechristoffel}
\eeq
to minimize the appearances of sign factors. On an antisymplectic manifold
\mb{(M;E)}, it is furthermore possible to define a Christoffel symbol with
three lower indices
\beq
\Gamma_{ABC}~\equiv~E_{AD}\Gamma^{D}{}_{BC}(-1)^{\eps_{B}}~.
\label{lowerchristoffel}
\eeq 
Let us also define
\beq
\gamma_{ABC}^{}~\equiv~\Gamma_{ABC}^{}-\frac{1}{3}(E_{A\{B}^{}\rpa{C\}})
~\equiv~\Gamma_{ABC}^{}-\frac{1}{3}(E_{AB}^{}\rpa{C}
+E_{AC}^{}\rpa{B}(-1)^{\eps_{B}\eps_{C}})~.
\label{christoffeltensor}
\eeq
\mb{\gamma_{ABC}} is {\em not} a tensor but it still has some useful
properties, see \eqs{smallgammaantisympl}{torsionfree3} below. One can think of
\mb{\gamma_{ABC}} as parametrizing all the possible connections \mb{\nabla} on
\mb{(M;E)}. 

\noi
An {\em antisymplectic connection} \mb{\Gamma_{A}{}^{B}{}_{C}} satisfies by
definition \cite{b97}
\beq
0~=~(\nabla_{A}E)^{BC}
~\equiv~(\lpa{A}E^{BC})+\left(\Gamma_{A}{}^{B}{}_{D}E^{DC}
-(-1)^{(\eps_{B}+1)(\eps_{C}+1)}(B\leftrightarrow C)\right)~,
\label{connuppereab}
\eeq
so that the antisymplectic metric \mb{E^{AB}} is covariantly preserved.  
In terms of the two-form \mb{E_{AB}}, the antisymplectic condition reads
\beq
0~=~(\nabla_{A} E)_{BC}
~\equiv~(\lpa{A}E_{BC})
-\left((-1)^{\eps_{A}\eps_{B}}\Gamma_{BAC}
-(-1)^{\eps_{B}\eps_{C}}(B\leftrightarrow C)\right)~.\label{connlowereab}
\eeq
Written in terms of the \mb{\gamma_{ABC}} symbol, the antisymplectic condition
\e{connlowereab} becomes a purely algebraic equation, due to the closeness
relation \e{abclose}:
\beq
\gamma_{ABC}
~=~(-1)^{\eps_{A}\eps_{B}+\eps_{B}\eps_{C}+\eps_{C}\eps_{A}}\gamma_{CBA}~.
\label{smallgammaantisympl}
\eeq

\noi
A {\em torsion-free} connection has the following symmetry in the lower
indices:
\bea
\Gamma^{A}{}_{BC}&=&-(-1)^{(\eps_{B}+1)(\eps_{C}+1)}\Gamma^{A}{}_{CB}~,
\label{torsionfree1} \\
\Gamma_{ABC}&=&(-1)^{\eps_{B}\eps_{C}}\Gamma_{ACB}~,\label{torsionfree2} \\
\gamma_{ABC}&=&(-1)^{\eps_{B}\eps_{C}}\gamma_{ACB}~.\label{torsionfree3}
\eea
Note that \mb{(-1)^{\eps_{A}\eps_{B}}\gamma_{BAC}=\gamma_{ABC}=
(-1)^{\eps_{B}\eps_{C}}\gamma_{ACB}} is totally symmetric for an antisymplectic
torsion-free connection. (Similar results hold for even symplectic structures.)

\noi
A connection \mb{\nabla} can be used to define a divergence of a Bosonic vector
field \mb{X^{A}} as 
\beq
\str(\nabla X)~\equiv~(-1)^{\eps_{A}}(\nabla_{A}X)^{A}~=~
((-1)^{\eps_{A}}\lpa{A}+\Gamma^{B}{}_{BA})X^{A}~,~~~~~~~~~\eps_{X}~=~0~.
\eeq
On the other hand, the divergence is defined in terms of \mb{F} or \mb{\rho} as
\bea
{\rm div}_{F}^{}X&\equiv&(-1)^{\eps_{A}}(\lpa{A}\!+\!F_{A}^{})X^{A}~,
\label{divf} \\
{\rm div}_{\rho}^{}X&\equiv&\frac{(-1)^{\eps_{A}}}{\rho} 
\lpa{A}(\rho X^{A})~. \label{divrho}
\eea
See \Ref{yks02} for a mathematical exposition of divergence operators on 
supermanifolds. Under the \mb{F\leftrightarrow \rho} identification \e{frho},
the last two Definitions \es{divf}{divrho} agree:
\beq
{\rm div}_{F}^{}X~=~{\rm div}_{\rho}^{}X~. 
\eeq
In order to have a unique divergence operator (and hence a unique notion 
of volume), it is necessary to impose the following compatibility condition
between \mb{F_{A}} and the Christoffel symbols \mb{\Gamma^{A}{}_{BC}}: 
\beq
\Gamma^{B}{}_{BA}~=~(-1)^{\eps_{A}}F_{A}~.
\label{connf}
\eeq
We shall only consider antisymplectic, torsion-free, and \mb{F}-compatible
connections \mb{\nabla}, \ie connections that satisfy the three conditions
\e{connuppereab}, \es{torsionfree1}{connf}. The first and third condition
ensure the compatibility with \mb{E} and \mb{F}, respectively. The second (the
torsion-free condition) guarantees compatibility with the closeness relation
\e{abclose}. It can be demonstrated that connections satisfying these three
conditions exist locally for \mb{N>1}, where \mb{2N} denotes the number of
antisymplectic variables \mb{\Gamma^{A}}, \mb{A=1,\ldots,2N}. (There are
counterexamples for \mb{N\!=\!1} where \mb{\nabla} need not exist.) {}For
connections satisfying the three conditions, the \mb{\DeltaF} operator can be
written on a manifestly covariant form
\beq
 \DeltaF~=~\frac{(-1)^{\eps_{A}}}{2}\nabla_{A}E^{AB}\nabla_{B}
~=~\frac{(-1)^{\eps_{B}}}{2}E^{BA}\nabla_{A}\nabla_{B}~.\label{covdeltaf} 
\eeq

\section{Curvature}
\label{seccurv}

\noi
The Riemann curvature tensor \mb{R_{AB}{}^{C}{}_{D}} is defined as
the commutator of the \mb{\nabla} connection
\beq
([\nabla_{A},\nabla_{B}]X)^{C}
~=~R_{AB}{}^{C}{}_{D}X^{D}(-1)^{\eps_{X}(\eps_{C}+\eps_{D})}~,
\eeq
so that
\beq
R_{AB}{}^{C}{}_{D}
~=~(\lpa{A}\Gamma_{B}{}^{C}{}_{D})
+(-1)^{\eps_{B}\eps_{C}}\Gamma_{A}{}^{C}{}_{E}\Gamma^{E}{}_{BD}
-(-1)^{\eps_{A}\eps_{B}}(A\leftrightarrow B)~. 
\eeq
It is useful to define a reordered Riemann curvature tensor
\mb{R^{A}{}_{BCD}} as
\beq
R^{A}{}_{BCD}~\equiv~(-1)^{\eps_{A}(\eps_{B}+\eps_{C})}R_{BC}{}^{A}{}_{D}
~=~(-1)^{\eps_{A}\eps_{B}}(\lpa{B}\Gamma^{A}{}_{CD})
+\Gamma^{A}{}_{BE}\Gamma^{E}{}_{CD}
-(-1)^{\eps_{B}\eps_{C}}(B\leftrightarrow C)~. 
\eeq
It is interesting to consider the various contractions of the Riemann curvature
tensor. There are two possibilities. {}Firstly,  there is the Ricci two-form 
\beq
\cR_{AB}~\equiv~R_{AB}{}^{C}{}_{C}(-1)^{\eps_{C}}~=~
(\lpa{A}F_{B})
-(-1)^{\eps_{A}\eps_{B}}(A \leftrightarrow B)~.\label{ricciform}
\eeq
However, the Ricci two-form \mb{\cR_{AB}} typically vanishes, \cf
\eq{zerocurvature}, and even if it does not vanish, its antisymmetry
\e{lowercrsymmetry} means that \mb{\cR_{AB}} cannot successfully be contracted
with the antisymplectic metric \mb{E^{AB}} to yield a non-zero scalar
curvature, \cf \eq{skewsym1}. Secondly, there is the Ricci tensor
\beq
R_{AB}~\equiv~R^{C}{}_{CAB} 
~=~(-1)^{\eps_{C}}(\lpa{C}+ F_{C})\Gamma^{C}{}_{AB}
-(\lpa{A}F_{B})(-1)^{\eps_{B}}
-\Gamma_{A}{}^{C}{}_{D}\Gamma^{D}{}_{CB}~.\label{riccitensor} 
\eeq
Note that when the torsion tensor and Ricci two-form vanish, the Ricci tensor
\mb{R_{AB}} possesses exactly the same \mb{A\!\leftrightarrow\!B} symmetry
\e{skewsym1} as the antisymplectic metric \mb{E^{AB}} with upper indices
\beq
R_{AB}~=~-(-1)^{(\eps_{A}+1)(\eps_{B}+1)}R_{BA}~. \label{rsymmetry}
\eeq
The {\em odd scalar curvature} \mb{R} is therefore defined in antisymplectic
geometry as the contraction of the Ricci tensor \mb{R_{AB}} and the
antisymplectic metric \mb{E^{BA}},
\beq
R~\equiv~R_{AB}E^{BA}~=~E^{AB}R_{BA}~.\label{defosc}
\eeq
 
\begin{proposition}
{}For an arbitrary, antisymplectic, torsion-free, and \mb{F}-compatible
connections \mb{\nabla}, the scalar curvature \mb{R} does only depend on
\mb{E} and \mb{F} through the odd scalar \mb{\nuF},
\beq
R~=~-8\nuF~,\label{rnuf}
\eeq
even if the line bundle connection \mb{F} is not flat.
\label{propositionc}
\end{proposition}

\noi
Proposition~\ref{propositionc} is shown in Appendix~\ref{apposc}. In
particular, one concludes that the scalar curvature \mb{R} does not depend on
the connection \mb{\Gamma^{A}{}_{BC}} used.  

\noi
One can perform various consistency checks on the formalism. Here, let us just
mention one. {}For an antisymplectic connection \mb{\nabla}, one has
\beq
0~=~[\nabla_{A},\nabla_{B}]E^{CD}~=~
R_{AB}{}^{C}{}_{F}E^{FD}
-(-1)^{(\eps_{C}+1)(\eps_{D}+1)}(C\leftrightarrow D)~,
\label{ddetjek1}
\eeq
or, equivalently,
\beq
R^{C}{}_{ABF}E^{FD}
~=~-(-1)^{\eps_{A}\eps_{B}+(\eps_{C}+1)(\eps_{D}+1)
+(\eps_{A}+\eps_{B})(\eps_{C}+\eps_{D})}R^{D}{}_{BAF}E^{FC}~.\label{ddetjek2}
\eeq
Contracting the \mb{A\leftrightarrow C} and \mb{B\leftrightarrow D} indices in
\eq{ddetjek2} indeed produces the identity \mb{R=R}. Had the signs turn out
differently, the odd scalar curvature \e{defosc} would have been stillborn, \ie
always zero.

\section{Conclusions}
\label{secconcl}

\noi
In this paper, we have first of all analyzed a general non-degenerate,
second-order \mb{\Delta} operator, and found that nilpotency determines the
\mb{\Delta} operator uniquely (after dismissing an odd constant). The result
is that \mb{\Delta} has to be \mb{\Deltarho\!+\!\nurho}, where \mb{\Deltarho}
is the odd Laplacian, and \mb{\nurho} is an odd scalar function (=zeroth-order
operator) that only depends on the density \mb{\rho} and the antisymplectic
structure \mb{E}. 
Secondly, we have shown that several constructions in antisymplectic geometry
can be extended to a non-flat line bundle connection \mb{F}, which replaces 
\mb{\rho}. We did this by breaking the nilpotency \mb{\Delta^{2}=\Hf\DeltaR}
by a general second-order operator \mb{\DeltaR}, which acts as a source for
the \mb{F}-curvature \mb{\cR}. In this more general case, the \mb{\Delta}
operator takes the form \mb{\DeltaF\!+\!\nuF\!+\!\nuR}, where \mb{\DeltaF} and
\mb{\nuF} are generalizations of the odd Laplacian \mb{\Deltarho} and the odd
scalar \mb{\nurho}, respectively. The \mb{\nuR} term is an auxiliary curvature
background encoded in the \mb{\DeltaR} operator. Thirdly, we have identified
the \mb{\nuF} function with (minus \mb{1/8} times) the odd scalar curvature
\mb{R} of an arbitrary antisymplectic, torsion-free, and \mb{F}-compatible
connection.

\noi
One may summarize by saying that two notions of curvature play an important
r\^ole in this paper:
1) a line bundle curvature \mb{\cR_{AB}} defined in \eq{crcurvature} and
2) an odd scalar curvature \mb{R} defined in \eq{defosc}. The former provides a
natural framework for several mathematical constructions, but it remains
currently unclear if it would be useful in physics. On the other
hand, the field-antifield formalism naturally embraces the latter type of
curvature both physically and mathematically. Concretely, we saw that the odd
scalar curvature \mb{R} manifests itself via a zeroth-order term \mb{\nuF} in
the \mb{\Delta} operator, which could potentially be used in a physical
application some day. Altogether, the odd scalar curvature \mb{R} and \mb{\nuF}
represent an important milestone in our understanding of the symmetries and the
supergeometric structures behind the powerful field-antifield formalism.

\vspace{0.8cm}

\noi
{\sc Acknowledgements:}~We would like to thank P.H.~Damgaard for discussions
and the Niels Bohr institute for warm hospitality. The work of I.A.B.\ is
supported by grants RFBR 05-01-00996, RFBR 05-02-17217 and LSS-4401.2006.2. The
work of K.B.\ is supported by the Ministry of Education of the Czech Republic
under the project MSM 0021622409.

\appendix

\section{Independence of Gauge-Fixing in the \mb{F}-Independent Formalism}
\label{appindep}

\noi
In this Appendix~\ref{appindep}, we prove in two different ways that the
partition function \e{epartitionfct} is independent of gauge-fixing. Let us
introduce the following shorthand notation
\beq
w~\equiv~e^{\Ih\WE}~,~~~~~~~~~x~\equiv~e^{\Ih\XE}~,
\eeq
for the two Boltzmann semidensities, so that the partition function
\e{epartitionfct} simply becomes
\beq
{\cal Z}~=~\int\! [d\Gamma][d\lambda]~w~x~. \label{epartitionfctshorthand}
\eeq  
The Boltzmann semidensities \mb{w} and \mb{x} are \mb{\DeltaE}-closed because
of the two quantum master eqs.~\e{eqmewx}. Since the \mb{\DeltaE} operator is
nilpotent, one may argue on general grounds that an arbitrary infinitesimal
variation of \mb{x} should be \mb{\DeltaE}-exact, which may be written as
\beq
\delta x~=~[\stackrel{\rightarrow}{\Delta}_{E},\delta\Psi]x
~\equiv~\DeltaE(\delta\Psi~x)+\delta\Psi (\DeltaE x)~,
\label{expmaxvar}
\eeq
if one assumes that \mb{x} is invertible and satisfies the quantum master
\eq{eqmewx}. Phrased equivalently, the variation \mb{\delta\XE} of the quantum
action is BRST-exact,
\beq
\delta\XE~=~(\XE,\delta\Psi)+\Hi\Deltaone(\delta\Psi)
~=~\sigma_{\XE}^{}(\delta\Psi)~,\label{maxvar}
\eeq
where \mb{\sigma_{\XE}^{}\!=\!(\XE,\cdot)+\Hi\Deltaone} is a quantum
BRST-operator. One may now proceed in at least two ways. One axiomatic way
\cite{bbd96} uses that the \mb{\DeltaE} operator \e{deltaedef} is symmetric,
\beq
\Delta_{E}^{T}~=~\DeltaE~,\label{deltasym}
\eeq
\ie stabile under integration by part. Then, an infinitesimal variation
\e{expmaxvar} of the gauge-fixing Boltzmann semidensity \mb{x} changes the
partition function as 
\bea
\delta{\cal Z}&=&\int\! [d\Gamma][d\lambda]~w~\delta x
~=~\int\![d\Gamma][d\lambda]~
w~[\stackrel{\rightarrow}{\Delta}_{E},\delta \Psi]x \cr
&=&\int\![d\Gamma][d\lambda]\left[(\DeltaE w)~\delta \Psi~x+
w~\delta\Psi~(\DeltaE x) \right]~=~0~,
\eea
where the symmetry property \e{deltasym} is used in the third equality and the
two quantum master equations \e{eqmewx} in the fourth (= last) equality.
Notice how this proof requires very little knowledge of the detailed form of
\mb{\DeltaE}. Another proof \cite{bt93,bv81,bms95} uses an intrinsic
infinitesimal redefinition of the integration variables,
\beq
\delta\Gamma^{A}
~=~\frac{i}{2\hbar}(\Gamma^{A},\XE\!-\!\WE)\delta\Psi
+\Hf(\Gamma^{A},\delta \Psi)
~=~\frac{w}{2x}(\Gamma^{A},\frac{x~\delta \Psi}{w})
~,~~~~~~~~\delta\lambda^{\alpha}~=~0~,
\label{gammavariation}
\eeq
to induce the allowed variation \e{expmaxvar} of \mb{x}. Now it is instructive
to write the path integral integrand as a volume form
\mb{\Omega\equiv w x[d\Gamma][d\lambda]} with measure density \mb{w x}. The
Lie-derivative is
\beq
\delta\Omega~=~({\rm div}_{w x}\delta\Gamma)\Omega~.
\eeq
In detail, the field-antifield redefinition \e{gammavariation} yields the
following logarithmic variation of \mb{\Omega}:
\bea
{\rm div}_{w x}\delta\Gamma&\equiv&
\frac{(-1)^{\eps_{A}}}{w x} \lpa{A}(w x~\delta\Gamma^{A})
~=~\frac{(-1)^{\eps_{A}}}{2w x} 
\lpa{A} w^{2}(\Gamma^{A},\frac{x~\delta\Psi}{w})
~=~\frac{w}{x} \Delta_{w^{2}}\frac{x~\delta \Psi}{w} \cr
&=&\frac{1}{x} \Deltaone(x~\delta \Psi)
-(\Deltaone w)\frac{\delta\Psi}{w}
~=~\frac{1}{x} \Deltaone(x~\delta\Psi)
+(\frac{\nu^{(1)}}{8}-\frac{\nu^{(2)}}{24})\delta\Psi
~=~\frac{1}{x} \DeltaE(x~\delta \Psi)\cr
&=&\frac{1}{x}[\stackrel{\rightarrow}{\Delta}_{E},\delta\Psi]x~=~\delta\ln x~.
\label{dxfromdummy}
\eea
Here, a non-trivial property of the odd Laplacian \e{deltarho} is used in the
fourth equality, the two quantum master equations \e{eqmewx} are used in the
fifth and seventh equality, and the formula \e{expmaxvar} for the allowed
variation of \mb{x} is used in the eighth (=last) equality. If one reads the
above \eq{dxfromdummy} in the opposite direction, one sees that all allowed
variations \e{expmaxvar} of the gauge-fixing Boltzmann semidensity \mb{x} can
be reproduced by an intrinsic field-antifield redefinition \e{gammavariation},
\beq
\delta{\cal Z}~=~\int\![d\Gamma][d\lambda]~w~\delta x~=~
\int\!\Omega~\delta\ln x~=~
\int\!\Omega~{\rm div}_{w x}\delta\Gamma
~=~\int\!\delta\Omega~=~0~.
\eeq
One concludes that the partition function \mb{{\cal Z}=\int\!\Omega} must be
independent of the gauge-fixing \mb{x} part since an intrinsic redefinition of
dummy integration variables cannot change the value of the path integral.

\section{Proof of Proposition~\ref{propositiona}}
\label{appnuf} 

\noi
In this Appendix~\ref{appnuf}, we show that the \mb{\nuF} expression \e{nuf}
satisfies the differential \eq{nueq} for the difference \mb{\nu\!-\!\nuR}. We
start by recalling that the \mb{\DeltaF} operator \e{deltaf} is
\beq
\DeltaF~\equiv~\Deltaone+V~,\label{deltafdelta1v}
\eeq
where \mb{\Deltaone} denotes the expression \e{deltarho} for the odd Laplacian
\mb{\Delta_{\rho=1}^{}} with \mb{\rho} replaced by \mb{1}, and where we, for
convenience, have defined
\beq
V~\equiv~\frac{(-1)^{\eps_{A}}}{2}F_{A}(\Gamma^{A},\cdot)~.
\eeq

\begin{lemma}
The square of the \mb{\DeltaF} operator is
\beq
\Delta_{F}^{2}~\equiv~\Delta_{1}^{2}+[\Deltaone,V]+V^{2}
~=~\Delta_{1}^{2}+\Hf\DeltaFR+(\nu_{F}^{(0)},\cdot)~. \label{extractingnuf0}
\eeq
\label{lemmab}
\end{lemma}

\noi
{\sc Proof of Lemma~\ref{lemmab}:}~~One finds by straightforward calculations
that
\bea
4V^2&=&(-1)^{\eps_{A}+\eps_{B}}F_{A}(\Gamma^{A},F_{B}(\Gamma^{B},\cdot)) \cr
&=&(-1)^{\eps_{A}}F_{B}F_{A}(\Gamma^{A},(\Gamma^{B},\cdot))
+(-1)^{\eps_{A}+\eps_{B}}F_{A}E^{AC}(\lpa{C}F_{B})(\Gamma^{B},\cdot) \cr
&=&\frac{(-1)^{\eps_{A}}}{2}F_{B}F_{A}((\Gamma^{A},\Gamma^{B}),\cdot)
+(-1)^{\eps_{A}}F_{A}E^{AC}
 [F_{C}\rpa{B}+\cR_{CB}(-1)^{\eps_{B}}](\Gamma^{B},\cdot)  \cr
&=&\frac{(-1)^{\eps_{A}}}{2}(F_{A}E^{AB}F_{B},\cdot)
+(-1)^{\eps_{A}+\eps_{C}}F_{A}E^{AB}\cR_{BC}(\Gamma^{C},\cdot)~,
\label{extractingnuf0a}
\eea
and
\bea
2[\Deltaone,V]&=&(-1)^{\eps_{A}}\Deltaone F_{A}(\Gamma^{A},\cdot)
+(-1)^{\eps_{A}}F_{A}(\Gamma^{A},\Deltaone(\cdot)) \cr
&=&(-1)^{\eps_{A}}(\Deltaone F_{A})(\Gamma^{A},\cdot)
+(F_{A},(\Gamma^{A},\cdot))
+F_{A}\Deltaone(\Gamma^{A},\cdot)
+(-1)^{\eps_{A}}F_{A}(\Gamma^{A},\Deltaone(\cdot)) \cr
&=&\frac{(-1)^{\eps_{A}+\eps_{B}}}{2}
(\lpa{B}E^{BC}\lpa{C}F_{A})(\Gamma^{A},\cdot)
+(F_{A}\rpa{B})(\Gamma^{B},(\Gamma^{A},\cdot))
+F_{A}(\Deltaone\Gamma^{A},\cdot) \cr
&=&\frac{(-1)^{\eps_{B}}}{2}
(\lpa{B}E^{BC}[F_{C}\rpa{A}+\cR_{CA}(-1)^{\eps_{A}}])(\Gamma^{A},\cdot) \cr
&&+\Hf[F_{A}\rpa{B}+(-1)^{\eps_{B}}\lpa{A}F_{B}
+(-1)^{(\eps_{A}+1)\eps_{B}}\cR_{BA}](\Gamma^{B},(\Gamma^{A},\cdot))
+F_{A}(\Deltaone\Gamma^{A},\cdot) \cr
&=&\frac{(-1)^{\eps_{C}}}{2}E^{CB}(\lpa{B}F_{C},\cdot)
+(\Deltaone\Gamma^{C})(F_{C},\cdot)
+\frac{(-1)^{\eps_{A}+\eps_{B}}}{2}
(\lpa{B}E^{BC}\cR_{CA})(\Gamma^{A},\cdot) \cr
&&+\frac{(-1)^{\eps_{B}}}{2}(\lpa{A}F_{B})((\Gamma^{B},\Gamma^{A}),\cdot)
-\frac{(-1)^{(\eps_{A}+1)(\eps_{C}+1)}}{2}
E^{CB}\cR_{BA}\lpa{C}(\Gamma^{A},\cdot) 
+F_{A}(\Deltaone\Gamma^{A},\cdot) \cr
&=&\frac{(-1)^{\eps_{B}}}{2}(E^{BA}\lpa{A}F_{B},\cdot)
+(F_{A}\Deltaone\Gamma^{A},\cdot)
+\frac{(-1)^{\eps_{A}+\eps_{C}}}{2}\lpa{A}E^{AB}\cR_{BC}(\Gamma^{C},\cdot) \cr
&=&\frac{(-1)^{\eps_{A}}}{2}(\lpa{A}(E^{AB}F_{B}),\cdot)
+\frac{(-1)^{\eps_{A}+\eps_{C}}}{2}\lpa{A}E^{AB}\cR_{BC}(\Gamma^{C},\cdot)~,
\label{extractingnuf0b}
\eea
where the Jacobi identity \e{jacid} has been applied in the third and fifth 
equality of \eqs{extractingnuf0a}{extractingnuf0b}, respectively. 
\proofbox

\noi
(As an aside, we mention that Lemma~\ref{lemmab} can be used to prove
Lemma~\ref{lemmaa} in Section~\ref{secdnc}.) When one compares
Lemma~\ref{lemmab} with the \mb{\nu} differential \eq{nueq}, one sees the first
clue that the \mb{\nuF} expression \e{nuf} is a solution. More precisely, 
Lemma~\ref{lemmab} has extracted the \mb{\nu_{F}^{(0)}} part for us. Next task
is to uncover the \mb{\nu^{(1)}} term \e{nu1}.

\begin{lemma}
\beq
8(\Delta_{1}^{2}\Gamma^{A})~=~(\nu^{(1)},\Gamma^{A})
-(-1)^{\eps_{C}}(\lpa{B}E^{CD})(\lpa{D}\lpa{C}E^{BA})~.\label{extractingnu1} 
\eeq
\label{lemmac}
\end{lemma}

\noi
{\sc Proof of Lemma~\ref{lemmac}:}~~Combine
\beq
(\lpa{B}\Deltaone E^{BA})-2(\Delta_{1}^{2}\Gamma^{A})
~=~[\lpa{B},\Deltaone]E^{BA}
~=~\Hf(-1)^{\eps_{C}}(\lpa{B}E^{CD})(\lpa{D}\lpa{C}E^{BA})
+(\lpa{B}\Deltaone\Gamma^{C})\lpa{C}E^{BA}~,
\eeq
and
\bea
(\lpa{B}\Deltaone E^{BA})
&=&\lpa{B}\Deltaone(\Gamma^{B},\Gamma^{A})
~=~\lpa{B}(\Deltaone\Gamma^{B},\Gamma^{A})
-(-1)^{\eps_{B}}\lpa{B}(\Gamma^{B},\Deltaone\Gamma^{A}) \cr
&=&\Hf (\nu^{(1)},\Gamma^{A})
+(\lpa{C}\Deltaone\Gamma^{B})(\lpa{B}E^{CA})
-2(\Delta_{1}^{2}\Gamma^{A})~.
\eea
\proofbox

\noi
So far, we have reproduced the \mb{\nu_{F}^{(0)}} and the \mb{\nu^{(1)}} part
of the \mb{\nuF} solution to the \mb{\nu} differential \eq{nueq}.
{}Finally, we should extract the \mb{\nu^{(2)}} term \e{nu2}. The prefactor 
\mb{1/24} in the \mb{\nuF} formula \e{nuf} hints that such a calculation is
going be lengthy. Rewrite first Lemma~\ref{lemmac} as
\beq
8(\Delta_{1}^{2}\Gamma^{B})E_{BA}~=~(\lpa{A}\nu^{(1)})-\nu_{A}^{I}~, 
\label{extractingnu1prime}  
\eeq
where
\bea
\nu_{A}^{I}&\equiv&
(-1)^{\eps_{D}}(\lpa{C}E^{DF})(\lpa{F}\lpa{D}E^{CB})E_{BA}
~=~\nu_{A}^{II}+\nu_{A}^{III}~, \label{nua1} \\
\nu_{A}^{II}&\equiv&(-1)^{\eps_{B}\eps_{D}}
(\lpa{D}E^{BC})(\lpa{C}E^{DF})(\lpa{F}E_{BA})
~=~-\nu_{A}^{II}-\nu_{A}^{IV}~, \label{nua2} \\
\nu_{A}^{III}&\equiv&
(-1)^{\eps_{D}}(\lpa{C}E^{DF})\lpa{F}((\lpa{D}E^{CB})E_{BA})  \cr
&=&-(-1)^{(\eps_{B}+\eps_{C})\eps_{D}}
(\lpa{C}E^{DF})\lpa{F}(E^{CB}\lpa{D}E_{BA})
~=~\nu_{A}^{II}+\nu_{A}^{V}~, \label{nua3} \\
\nu_{A}^{IV}&\equiv&(-1)^{\eps_{C}\eps_{D}}
(\lpa{A}E_{BC})(\lpa{D}E^{CF})(\lpa{F}E^{DB})~, \label{nua4} \\
\nu_{A}^{V}&\equiv&(-1)^{\eps_{C}}E^{BF}(\lpa{F}E^{CD})(\lpa{D}\lpa{C}E_{BA})
~=~-2\nu_{A}^{VI}~, \label{nua5} \\
\nu_{A}^{VI}&\equiv&(-1)^{\eps_{B}(\eps_{C}+1)}
E^{CF}(\lpa{F}E^{BD})(\lpa{D}\lpa{C}E_{BA})
~=~\nu_{A}^{V}+\nu_{A}^{VII}~,  \label{nua6}\\
\nu_{A}^{VII}&\equiv&(-1)^{\eps_{C}}
(\lpa{A}\lpa{B}E_{CD})E^{DF}(\lpa{F}E^{CB})~. \label{nua7} 
\eea
Here, the Jacobi identity \e{abjacid} is used in the second equality of
\eq{nua5}, and the closeness relation \e{abclose} is used in the second
equalities of \eqs{nua2}{nua6}. Altogether eqs.\ \e{nua1}--\e{nua7} yield
\beq
\nu_{A}^{I}~=~\nu_{A}^{II}+\nu_{A}^{III}~=~2\nu_{A}^{II}+\nu_{A}^{V}
~=~-\nu_{A}^{IV}+\nu_{A}^{V}~=~-\nu_{A}^{IV}-\frac{2}{3}\nu_{A}^{VII}~.
\label{nua1prime}
\eeq
Ultimately, we would like to show that \mb{\nu_{A}^{I}} is equal to
\mb{(\lpa{A}\nu^{(2)})/3}. The achievement in \eq{nua1prime} is more modest:
The free ``\mb{A}'' index on the \mb{\nu_{A}^{I}} expression has been moved to
a derivative \mb{\lpa{A}} in \mb{\nu_{A}^{IV}} and \mb{\nu_{A}^{VII}}. 
On the other hand, differentiation \wrt \mb{\Gamma^{A}} of the two expressions
\es{nu2}{mnu3} for the \mb{\nu^{(2)}} quantity \e{nu2} yields two more
relations
\beq
\nu_{A}^{IV}+2\nu_{A}^{VIII}~=~(\lpa{A}\nu^{(2)})~=~
\nu_{A}^{VIII}-\nu_{A}^{VII}-\nu_{A}^{IX}~, \label{danu2}
\eeq
where
\bea
\nu_{A}^{VIII}&\equiv&(-1)^{\eps_{C}\eps_{F}}
(\lpa{A}\lpa{B}E^{CD})E_{DF}(\lpa{C}E^{FB})~, \label{nua8} \\
\nu_{A}^{IX}&\equiv&(-1)^{\eps_{C}}
(\lpa{A}E^{DF})(\lpa{F}E^{CB})(\lpa{B}E_{CD}) \cr
&=&-(-1)^{\eps_{B}\eps_{G}}(\lpa{A}E^{DF})
(\lpa{F}E^{BC})E_{CG}(\lpa{B}E^{GH})E_{HD} \cr
&=&(-1)^{\eps_{B}\eps_{G}}(\lpa{A}E_{HD})E^{DF}
(\lpa{F}E^{BC})E_{CG}(\lpa{B}E^{GH})
~=~\nu_{A}^{IV}-\nu_{A}^{X}~, \label{nua9} \\
\nu_{A}^{X}&\equiv&(-1)^{\eps_{B}\eps_{G}+(\eps_{B}+\eps_{C})(\eps_{D}+1)}
(\lpa{A}E_{HD})E^{BF}(\lpa{F}E^{CD})E_{CG}(\lpa{B}E^{GH}) \cr
&=&(-1)^{(\eps_{B}+1)\eps_{D}+\eps_{C}(\eps_{B}+\eps_{H}+1)}
(\lpa{A}E_{HD})E^{BF}(\lpa{F}E^{DC})(\lpa{B}E^{HG})E_{GC}~=~0~. \label{nua10}
\eea
Here, the Jacobi identity \e{jacid} is used in the fourth equality of
\eq{nua9}. Remarkably, the \mb{\nu_{A}^{X}} term vanishes due to an
antisymmetry under the index permutation \mb{FDC\leftrightarrow BHG}.
Altogether, \mb{\nu_{A}^{IX}=\nu_{A}^{IV}} and
\beq
\nu_{A}^{I}~=~-\nu_{A}^{IV}-\frac{2}{3}\nu_{A}^{VII}
~=~-\nu_{A}^{IV}-\frac{2}{3}(\nu_{A}^{VIII}-\nu_{A}^{IV}-\lpa{A}\nu^{(2)})
~=~\frac{1}{3}(\lpa{A}\nu^{(2)})~.
\label{nua1primeprime}
\eeq
Combining eqs.\ \e{extractingnuf0}, \es{extractingnu1prime}{nua1primeprime}
shows that the \mb{\nuF} expression \e{nuf} satisfies the \mb{\nu} differential
\eq{nueq}.

\section{Proof of Proposition~\ref{propositionc}}
\label{apposc} 

\noi
In this Appendix~\ref{apposc}, we prove that the odd scalar curvature \mb{R} is
minus eight times the odd scalar \mb{\nuF}. The odd scalar curvature
\beq
 R~\equiv~R_{AB}E^{BA}~=~R_{I}+R_{II}-R_{III}-R_{IV} \label{r1234}
\eeq
inherits four terms \mb{R_{I}}, \mb{R_{II}}, \mb{R_{III}} and \mb{R_{IV}} 
from the expression \e{riccitensor} for the Ricci tensor \mb{R_{AB}}.
They are defined as
\bea
R_{I}&\equiv&(-1)^{\eps_{A}}(\lpa{A}\Gamma^{A}{}_{BC})E^{CB}
~=~R_{V}-R_{VI}~,\label{r1}\\
R_{II}&\equiv&(-1)^{\eps_{A}}F_{A}\Gamma^{A}{}_{BC}E^{CB}
~=~-(-1)^{\eps_{B}}F_{A}(\lpa{B}+F_{B})E^{BA}~,\label{r2}\\
R_{III}&\equiv&(-1)^{\eps_{B}}E^{BA}(\lpa{A}F_{B})~,\label{r3}\\
R_{IV}&\equiv&\Gamma_{A}{}^{C}{}_{D}\Gamma^{D}{}_{CB}E^{BA}
~=~-R_{IV}-R_{VI}~,\label{r4}\\
R_{V}&\equiv&(-1)^{\eps_{A}}\lpa{A}(\Gamma^{A}{}_{BC}E^{CB})
~=~-(-1)^{\eps_{B}}\lpa{A}(\lpa{B}+F_{B})E^{BA} \cr
&=&-\nu^{(1)}-(-1)^{\eps_{A}}\lpa{A}(E^{AB}F_{B})~,\label{r5}\\
R_{VI}&\equiv&\Gamma^{A}{}_{BC}(E^{CB}\rpa{A})~.\label{r6}
\eea
Here, the antisymplectic and the torsion-free conditions
\es{connuppereab}{torsionfree1} are used in the second equality of \eq{r4}, and
a contracted version of the antisymplectic condition \e{connuppereab}
\beq
(-1)^{\eps_{B}}(\lpa{B}\!+\!F_{B}^{})E^{BA}
+(-1)^{\eps_{A}}\Gamma^{A}{}_{BC}E^{CB}~=~0
\eeq 
is used in the second equalities of \eqs{r2}{r5}.
Inserting back in \eq{r1234}, one finds that
\beq
R~=~-8\nu_{F}^{(0)}-\nu^{(1)}-\Hf R_{VI}~,\label{earlyrnuf}
\eeq
where \mb{\nu_{F}^{(0)}} and \mb{\nu^{(1)}} are given in \eqs{nuf0}{nu1}. Now
it remains to eliminate \mb{R_{VI}} from \eq{earlyrnuf}. Note that \mb{R_{VI}}
only depends on the torsion-free part of the connection \mb{\Gamma^{A}{}_{BC}},
so one does in principle not need the torsion-free condition \e{torsionfree1}
from now on. One calculates that
\bea
\Hf R_{VI}&=&
-\Hf(-1)^{\eps_{A}(\eps_{D}+1)}
\Gamma_{B}{}^{A}{}_{C}E^{CD}(\lpa{A}E_{DF})E^{FB}
~=~-(-1)^{\eps_{A}}\Gamma_{B}{}^{A}{}_{C}E^{CD}(\lpa{D}E_{AF})E^{FB} \cr
&=&\Gamma^{A}{}_{BC}E^{CD}(\lpa{D}E^{BF})E_{FA} 
~=~-\nu^{(2)}-\mb{R_{VI}}~.
\eea
Here, the closeness relation \e{abclose} is used in the second equality and the
antisymplectic condition \e{connuppereab} in the fourth equality. In other
words,
\beq
\mb{R_{VI}}~=~-\frac{2}{3}\nu^{(2)}~.\label{twothird}
\eeq
Combining \eqs{earlyrnuf}{twothird} yields the main result of
Proposition~\ref{propositionc}:
\beq
R~=~-8\nuF~.\label{rnufextra}
\eeq

\end{document}